\documentclass[usenatbib,useAMS]{mn2e}
\usepackage{graphics}
\usepackage{psfrag}
\usepackage{xspace}
\usepackage{amstext}
\usepackage{amsmath}
\usepackage{amssymb}
\usepackage{natbib}
\usepackage{aas_macros}
\usepackage{ulem}
\usepackage{graphicx}
\usepackage{longtable}
\usepackage{rotating}
\usepackage{float}
\usepackage{caption}
\usepackage{subcaption}
\usepackage{varwidth}

\newcommand{\FigDir}[1]{#1}

\newcommand{\Msun}{\ensuremath{M_{\odot}}\xspace}

\newcommand{\Mjup}{\ensuremath{M_{\rm J}}\xspace}

\newcommand{\Teff}{\ensuremath{T_{\rm eff}}\xspace}

\newcommand{\dl}{\ensuremath{D_{\rm L}}\xspace}
\newcommand{\ds}{\ensuremath{D_{\rm S}}\xspace}

\newcommand{\thE}{\ensuremath{\theta_{\rm E}}\xspace}
\newcommand{\thS}{\ensuremath{\theta_{\star}}\xspace}

\newcommand{\umin}{\ensuremath{u_{0}}\xspace}
\newcommand{\tzero}{\ensuremath{t_{0}}\xspace}
\newcommand{\tE}{\ensuremath{t_{\rm E}}\xspace}
\newcommand{\Ml}{\ensuremath{M_{\rm L}}\xspace}
\newcommand{\Mp}{\ensuremath{M_{\rm p}}\xspace}

\newcommand{\mas}{{\rm mas}\xspace}

\newcommand{\kpc}{\ensuremath{\rm kpc}\xspace}

\newcommand{\au}{{\rm AU}\xspace}

\newcommand{\enp}{OGLE-2014-BLG-0676Lb\xspace}
\newcommand{\enogle}{OGLE-2014-BLG-0676\xspace}
\newcommand{\enmoa}{MOA-2014-BLG-175\xspace}
\newcommand{\en}{\enogle/\enmoa}

\newlength{\voff}
\newcommand{\planetmassdashNoRem}{\ensuremath{3.09_{-1.12}^{+1.02}}\,\Mjup}
\newcommand{\planetorbitdashNoRem}{\ensuremath{4.40_{-1.46}^{+2.16}}\,\au}
\newcommand{\lensmassdashNoRem}{\ensuremath{0.62_{-0.22}^{+0.20}}\,\Msun}
\newcommand{\lensdistancedashNoRem}{\ensuremath{2.22_{-0.83}^{+0.96}}\,\kpc}

\newcommand{\Rmoa}{\ensuremath{R_{\rm MOA}}\xspace}
\newcommand{\IogleIII}{\ensuremath{I_{\rm OGLE-III}}\xspace}
\newcommand{\IogleIV}{\ensuremath{I_{\rm OGLE-IV}}\xspace}
\newcommand{\VmIogleIII}{\ensuremath{(V-I)_{\rm OGLE-III}}\xspace}
\newcommand{\VmIogleIV}{\ensuremath{(V-I)_{\rm OGLE-IV}}\xspace}

\begin{document}

\title[\enp: A gas giant planet]{Faint source star planetary microlensing: the discovery of the cold gas giant planet \enp}

\author[N.~J.~Rattenbury et al.]
{\parbox{\textwidth}{N.~J.~Rattenbury,$^{1}$$^{,A}$\thanks{E-mail: \texttt{n.rattenbury@auckland.ac.nz}}
D.~P.~Bennett,$^{2}$$^{,A}$
T.~Sumi,$^{3}$$^{,A}$
N.~Koshimoto,$^{3}$$^{,A}$
I.~A.~Bond,$^{4}$$^{,A}$
A.~Udalski,$^{5}$$^{,B}$
Y.~Shvartzvald,$^{6}$$^{,C}$
D.~Maoz,$^{7}$$^{,C}$
U.~G.~J\o{}rgensen,$^{8}$$^{,D}$
M.~Dominik,$^{9}$$^{,D,E}$
R.~A.~Street,$^{10}$$^{,E}$
Y.~Tsapras,$^{10,11}$$^{,E}$
F.~Abe,$^{12}$$^{,A}$
Y.~Asakura,$^{12}$$^{,A}$
R.~Barry,$^{13}$$^{,A}$
A.~Bhattacharya,$^{2}$$^{,A}$
M.~Donachie,$^{1}$$^{,A}$
P.~Evans,$^{1}$$^{,A}$
M.~Freeman,$^{14}$$^{,A}$
A.~Fukui,$^{15}$$^{,A}$
Y.~Hirao,$^{3}$$^{,A}$
Y.~Itow,$^{12}$$^{,A}$
M.~C.~A.~Li,$^{1}$$^{,A}$
C.~H.~Ling,$^{4}$$^{,A}$
K.~Masuda,$^{12}$$^{,A}$
Y.~Matsubara,$^{12}$$^{,A}$
Y.~Muraki,$^{12}$$^{,A}$
M.~Nagakane,$^{3}$$^{,A}$
K.~Ohnishi,$^{16}$$^{,A}$
H.~Oyokawa,$^{12}$$^{,A}$
To.~Saito,$^{17}$$^{,A}$
A.~Sharan,$^{1}$$^{,A}$
D.~J.~Sullivan,$^{18}$$^{,A}$
D.~Suzuki,$^{2}$$^{,A}$
P.~J.~Tristram,$^{19}$$^{,A}$
A.~Yonehara,$^{20}$$^{,A}$
R.~Poleski,$^{5,21}$$^{,B}$
J.~Skowron,$^{5}$$^{,B}$
P.~Mr{\'o}z,$^{5}$$^{,B}$
M.~K.~Szyma{\'n}ski,$^{5}$$^{,B}$
I.~Soszy{\'n}ski,$^{5}$$^{,B}$
P.~Pietrukowicz,$^{5}$$^{,B}$
S.~Koz{\l}owski,$^{5}$$^{,B}$
K.~Ulaczyk,$^{5,22}$$^{,B}$
{\L}.~Wyrzykowski,$^{5}$$^{,B}$
M.~Friedmann,$^{7}$$^{,C}$
S.~Kaspi,$^{7}$$^{,C}$
K.~Alsubai,$^{23}$$^{,D}$
P.~Browne,$^{9}$$^{,D}$
J.~M.~Andersen,$^{24}$$^{,D}$
V.~Bozza,$^{25,26}$$^{,D}$
S.~Calchi~Novati,$^{27,25}$$^{,D}$
Y.~Damerdji,$^{28}$$^{,D}$
C.~Diehl,$^{11}$$^{,D}$
S.~Dreizler,$^{29}$$^{,D}$
A.~Elyiv,$^{28}$$^{,D}$
E.~Giannini,$^{11}$$^{,D}$
S.~Hardis,$^{8}$$^{,D}$
K.~Harps\o{}e,$^{8}$$^{,D}$
T.~C.~Hinse,$^{30}$$^{,D}$
C.~Liebig,$^{9}$$^{,D}$
M.~Hundertmark,$^{11}$$^{,D,E}$
D.~Juncher,$^{8}$$^{,D}$
N.~Kains,$^{31,32}$$^{,D,E}$
E.~Kerins,$^{33}$$^{,D}$
H.~Korhonen,$^{8}$$^{,D}$
L.~Mancini,$^{34}$$^{,D}$
R.~Martin,$^{35}$$^{,D}$
M.~Mathiasen,$^{8}$$^{,D}$
M.~Rabus,$^{36,34}$$^{,D}$
S.~Rahvar,$^{37}$
G.~Scarpetta,$^{25,26,38}$$^{,D}$
J.~Skottfelt,$^{8}$$^{,D}$
C.~Snodgrass,$^{39}$$^{,D,E}$
J.~Surdej,$^{28}$$^{,D}$
J.~Taylor,$^{40}$$^{,D}$
J.~Tregloan-Reed,$^{40}$$^{,D}$
C.~Vilela,$^{40}$$^{,D}$
J.~Wambsganss,$^{11}$$^{,D,E}$
A.~Williams,$^{35}$$^{,D}$
G.~D'Ago,$^{25,38}$$^{,D,E}$
E.~Bachelet,$^{10}$$^{,E}$
D.~M.~Bramich,$^{23}$$^{,E}$
R.~Figuera Jaimes,$^{31}$$^{,E}$
K.~Horne,$^{9}$$^{,E}$
J.~Menzies,$^{41}$$^{,E}$
R.~Schmidt$^{11}$$^{,E}$ and
I.~A.~Steele$^{42}$$^{,E}$\\[20pt]{\it \small Affiliations appear at the end of the paper}}}

\date{Accepted ........
      Received .......;
      in original form ......}

\pubyear{2015}

\maketitle
\begin{abstract}
We report the discovery of a planet --- \enp --- via gravitational microlensing. Observations for the lensing event were made by the MOA, OGLE, Wise, RoboNET/LCOGT, MiNDSTEp and $\mu$FUN groups. All analyses of the light curve data favour a lens system comprising a planetary mass orbiting a host star. The most favoured binary lens model has a mass ratio between the two lens masses of $(4.78 \pm 0.13)\times 10^{-3}$. Subject to some important assumptions, a Bayesian probability density analysis suggests the lens system comprises a \planetmassdashNoRem planet orbiting a \lensmassdashNoRem host star at a deprojected orbital separation of \planetorbitdashNoRem. The distance to the lens system is \lensdistancedashNoRem. Planet \enp provides additional data to the growing number of cool planets discovered using gravitational microlensing against which planetary formation theories may be tested. Most of the light in the baseline of this event is expected to come from the lens and thus high-resolution imaging observations could confirm our planetary model interpretation.
\end{abstract}


\section{Introduction}
\label{sec:intro}
To date, 47 planets in 45 planetary systems have been discovered by  microlensing\footnote{exoplanet.eu}. Discovering planets via microlensing \citep{1991ApJ...374L..37M} offers a number of advantages over either the transit \citep{2011ApJ...736...19B} or radial velocity \citep{2005ApJ...619..570M} methods. Principal among these is the fact that gravitational microlensing is most sensitive to planets orbiting at a few AU away from typical host stars, which is of particular interest to planetary formation theorists. This is because  these orbital radii  correspond to the location of the ``snow-line''--- the radius around a star beyond which both volatile and refractory material is present in a protoplanetary disk and according to the core-accretion theory of planetary formation, just beyond the snow line is where the most active planet formation occurs \citep{2005ApJ...626.1045I}. Understanding the distribution of planets in this region is therefore of particular importance.

The probability that any given star is magnified by a foreground lens object is very small --- around $10^{-6}$ \citep{2013ApJ...778..150S}. For this reason, microlensing survey observations are carried out towards the dense stellar fields of the Galactic bulge and the Magellanic Clouds. The concomitant challenges of analysing crowded stellar field images, providing real-time analysis of light curve data and the timely notification of colleagues at instruments spaced in longitude regarding ongoing events have been tackled by the ground-based microlensing community over the last two decades. See \cite{2012ARA&A..50..411G} for a review.

Having demonstrated its worth as a planet-detection method, planet-hunters are now using space telescopes to obtain follow-up observations of gravitational microlensing events detected from ground-based surveys. This is a step towards the fully spaced-based observations proposed some time ago \citep{2002ApJ...574..985B, 2002ESASP.485..195R} and which now is being realised with  {\it Spitzer}  observations of microlensing events (see e.g. \cite{2016ApJ...823...63P, 2016ApJ...820...79B, 2015ApJ...802...76Y}). Furthermore, microlensing events are being monitored in Campaign 9 of the {\it K2} space telescope mission during the present 2016 microlensing season \citep{2015arXiv151209142H,2016arXiv160501059P}, and there is great anticipation regarding the performance of microlensing as part of either or both of the {\it Euclid} and {\it WFIRST} space telescope missions \citep{2013EPSC....8..837B, 2015arXiv150303757S,2014arXiv1409.2759Y}. 

For now, the work of the microlensing community remains --- in part --- to continue to report each planetary microlensing discovery, in order to power the statistical analyses that are informing us on the planet demographics of the Galaxy \citep{2016MNRAS.457.4089S, 2011Natur.473..349S,2012Natur.481..167C,2010ApJ...720.1073G}, and which may challenge the present understanding of how planetary systems form.

Several ground-based observation programmes routinely monitor dense stellar fields to search for microlensing events. The Microlensing Observations in Astrophysics collaboration (MOA, \cite{2001MNRAS.327..868B}, \cite{2003ApJ...591..204S}), the  Optical Gravitational Lensing Experiment (OGLE, \cite{2015AcA....65....1U}) and the Korea Microlensing Network (KMTNet, \cite{2016JKAS...49...37K}) are continuing their microlensing survey operations. More details on these collaborations are given in \cite{2015MNRAS.454..946R}.

The Wise Observatory Group operates a 1 metre telescope at Wise Observatory in Israel. This instrument comprises the Large Area Imager for the Wise Observatory (LAIWO) camera which has a 1 square degree field of view. At the time of event \en, the Wise Observatory was conducting a survey for microlensing events \citep{2016MNRAS.457.4089S}.

The survey collaborations issue alerts to the broader community when each new microlensing event is discovered. Particular attention is drawn to those events showing evidence for perturbations which could be owing to a planet, or which are predicted to have a high sensitivity to such perturbations. Confirmatory and supporting observations are sought and made by follow-up observation groups such as $\mu$-FUN \citep{2010ApJ...720.1073G}, PLANET \citep{1998ApJ...509..687A}, RoboNet \citep{2009AN....330....4T} and MiNDSTEp \citep{2010AN....331..671D}. Preliminary models explaining the data in hand are computed by human and artificial experts alike (see e.g. \cite{2010MNRAS.408.2188B,2012MNRAS.424..902B}\footnote{www.fisica.unisa.it/GravitationAstrophysics/RTModel/\ldots\allowbreak{}\phantom{XXXXXXXXXXXXXXXXXXXXXX}\ldots 2016/RTModel.htm}) to inform any further commitment of follow-up resources. 

This work reports the discovery of planet \enp via gravitational microlensing, using survey and follow-up data. In Sections~\ref{sec:observations} and \ref{sec:reduction} we describe the data and their treatment, respectively. We describe our modelling of the observed light curve data in Section~\ref{sec:modelling}. An analysis of the microlensed source star is given in Section~\ref{sec:sourcestar} and the planet parameters estimated from a probability density analysis are presented in Section~\ref{sec:likelihood}.  We discuss and conclude our work in Section~\ref{sec:conclusion}.

\section{Observations}

Microlensing event \enogle was discovered by the OGLE collaboration on 26 April 2014 at 21:19~UTC. The MOA collaboration discovered the same event 2.7519 days later, designating the event \enmoa. The OGLE coordinates of this event are (RA, Dec; J2000.0) = (17\wh52\wm24\fs50, -30\wdg32\wm54\fs20); $(l,b) = (359.376073^{\circ}, -2.092892^{\circ})$. 

The OGLE light curve for this event comprises 4115 data points in the I band. The MOA light curve comprises 24659 data points, taken in the custom MOA-red filter, which has a passband corresponding to the sum of the standard R and I filters. 468 I band observations were obtained with the Wise telescope. In addition to these survey data sets, further observations were made by the RoboNET collaboration using the Las Cumbres Observatory Global Telescope (LCGOT, \cite{2013PASP..125.1031B}), and the Danish 1.54 metre telescope at ESO La Silla, Chile as part of the MiNDSTEp microlensing follow-up programme \citep{2010AN....331..671D}. 7, 2 and 13 observations of the event in the SDSS-i passband were made using the 1 metre LCOGT telescopes at their SAAO, Siding Springs and CTIO observatories, respectively. 16 observations were made with the Danish 1.54 metre telescope.

\label{sec:observations}

\section{Data Reduction}
\label{sec:reduction}
The OGLE data for this event were generated from image data via the OGLE difference imaging pipeline \citep{2003AcA....53..291U}. The images obtained by the MOA telescope were reduced by the difference imagine pipeline of  \cite{2001MNRAS.327..868B}. When performing difference imaging on a crowded stellar field, the photometry of a star can be affected by a neighbouring star of a different colour, owing to the differential refraction effect of Earth's atmosphere. The MOA light curve data were corrected for differential refraction. The OGLE and Wise data did not require this correction. The Wise data were reduced using the pySis DIA software \citep{2009MNRAS.397.2099A}. The RoboNET data were reduced using DanDIA and the MiNDSTEp data were reduced using a modified version of DanDIA for use with EMCCDs \citep{2008MNRAS.386L..77B,2013MNRAS.428.2275B}. The MiNDSTEp instrumental R-band light curve data were converted to I-band magnitudes using a photometric calibration between the MiNDSTEp and OGLE photometry scales (M.~Hundertmark, private communication). The $\mu$-FUN collaboration took a few observations in the I and H bands, however these data were not obtained with a view to constraining the parameters of a binary light curve model and are therefore not included here. The MOA collaboration did not obtain any V-band images of this event.
While the OGLE collaboration did take some V-band images of the event field, the source star does not appear on their reference images,  thus no measurement of the magnified source is possible. 

The MOA and OGLE survey datasets were trimmed to include only data from the 2013 --- 2015 microlensing seasons for the purpose of finding candidate models to explain the data. This trimming was done to avoid any problems arising from any long timescale systematic errors in the event's baseline photometry, caused, for example, by flat-fielding inconsistencies over the timescales of years. The binary nature of the lens system is seen in the light curve data at times $JD' = JD-2450000 \in [6775,6779]$ days. Any data points outside this time interval which provide a contribution to $\chi^{2} >16$ with respect to a preliminary well-fitting binary lens model were discarded. Errors on the light curve data were rescaled to give $\chi^{2}/\mathrm{dof} =1$. The final data sets used for modelling are summarised in Table~\ref{tab:data}.

\begin{table}
\begin{center}
\caption{\label{tab:data}Final data sets used for modelling planetary microlensing event \en.}
\begin{tabular}{lll}
\hline
Group & Band & N \\
\hline
OGLE & I & 2530\\
MOA & MOA-Red & 9185\\
Wise & I & 444\\
LCOGT & SDSS-i & 22 \\
MiNDSTEp & R & 16\\
\hline
\end{tabular}
\end{center}
\end{table}

\section{Binary Lens Modelling}
\label{sec:modelling}

The parameters for the simplest binary lens model assuming a finte-sized source star are the mass ratio, $q= \Mp / \Ml $; the separation, $s$, of the two mass elements in the binary lens system;  the angle $\alpha$ that the path of the source star subtends with respect to the line defined by the two mass elements in the binary lens; the impact parameter \umin; the time interval \tE which the source takes to traverse the Einstein radius; the epoch \tzero which corresponds to when the source is at position \umin; and the source star radius $\rho \equiv \thS/\thE$. \thS and \thE are the angular sizes of the source star and Einstein ring radius respectively.  \umin is measured in units of \thE from the centre of mass of the lens system, $s$ is measured  in units of the Einstein ring radius and \Mp and \Ml are planet and host lens masses respectively.

 Initial binary lens models were found by performing a wide grid search over the parameters $q$, $s$ and $\alpha$, allowing the parameters \umin, \tzero, \tE and $\rho$ to vary. Starting from several good candidate binary lens solutions found in the grid search, we refined these models using an adaptation of the image-centred ray-shooting method of \cite{2010ApJ...716.1408B}. The limb-darkening of the source star is included in our modelling through a linear profile. The surface brightness of the source star is given by $S_\lambda (\theta) = S_\lambda (0)[1 - u_{\lambda}(1-\cos(\theta))]$ where $\theta$ is the angle between the line of sight and the normal to the stellar surface and $u_{\lambda}$ is the limb-darkening coefficient  for a particular filter colour. The source star's $(V-I)$ colour determines our values of the limb-darkening coefficients.  However, this colour estimate for the source is model dependent. We therefore iterated our analysis for producing a best-fitting binary lens model and the analysis leading to an estimate of the source colour --- and thence to values of $u_{\lambda}$, see Section~\ref{sec:sourcestar} --- until the iterative analysis converges on values of $u_{\lambda}$.

\begin{figure}
\includegraphics[width=84mm]{\FigDir{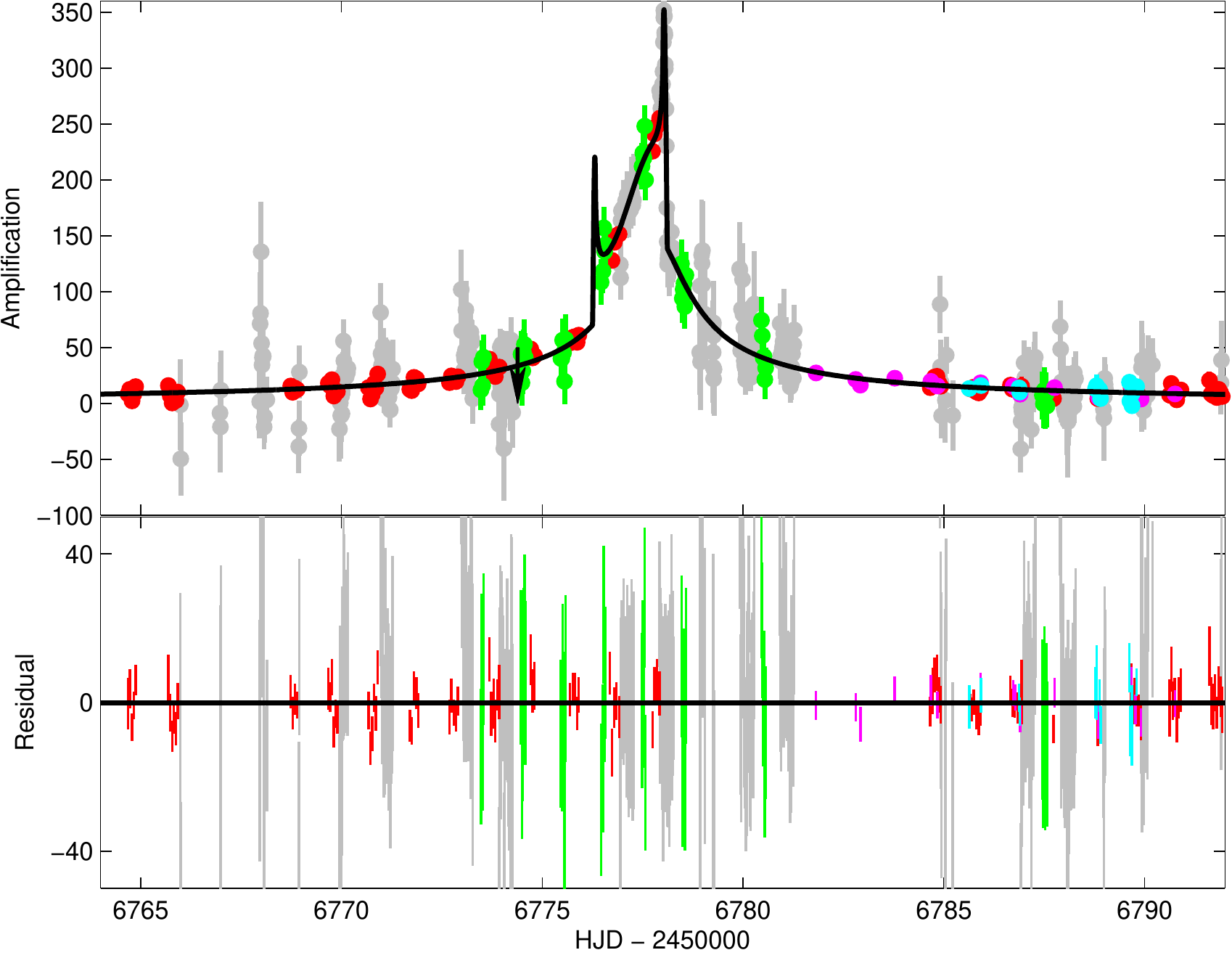}}
 \caption{\label{fig:lightcurve1}Observed data for event \en from the MOA (gray), OGLE (red)  and Wise (green) microlensing survey groups along with data from the RoboNET/LCOGT (cyan) and MiNDSTEp (magenta) groups.  Also shown is the best-fitting binary lens model light-curve (black line). The parameters for this model are given in Table~\ref{tab:parameters}. The epoch when the OGLE collaboration issued an alert for this event is indicated with a black arrow. Data with extremely large errors are omitted.}
  \end{figure}

\begin{figure}
\includegraphics[width=84mm]{\FigDir{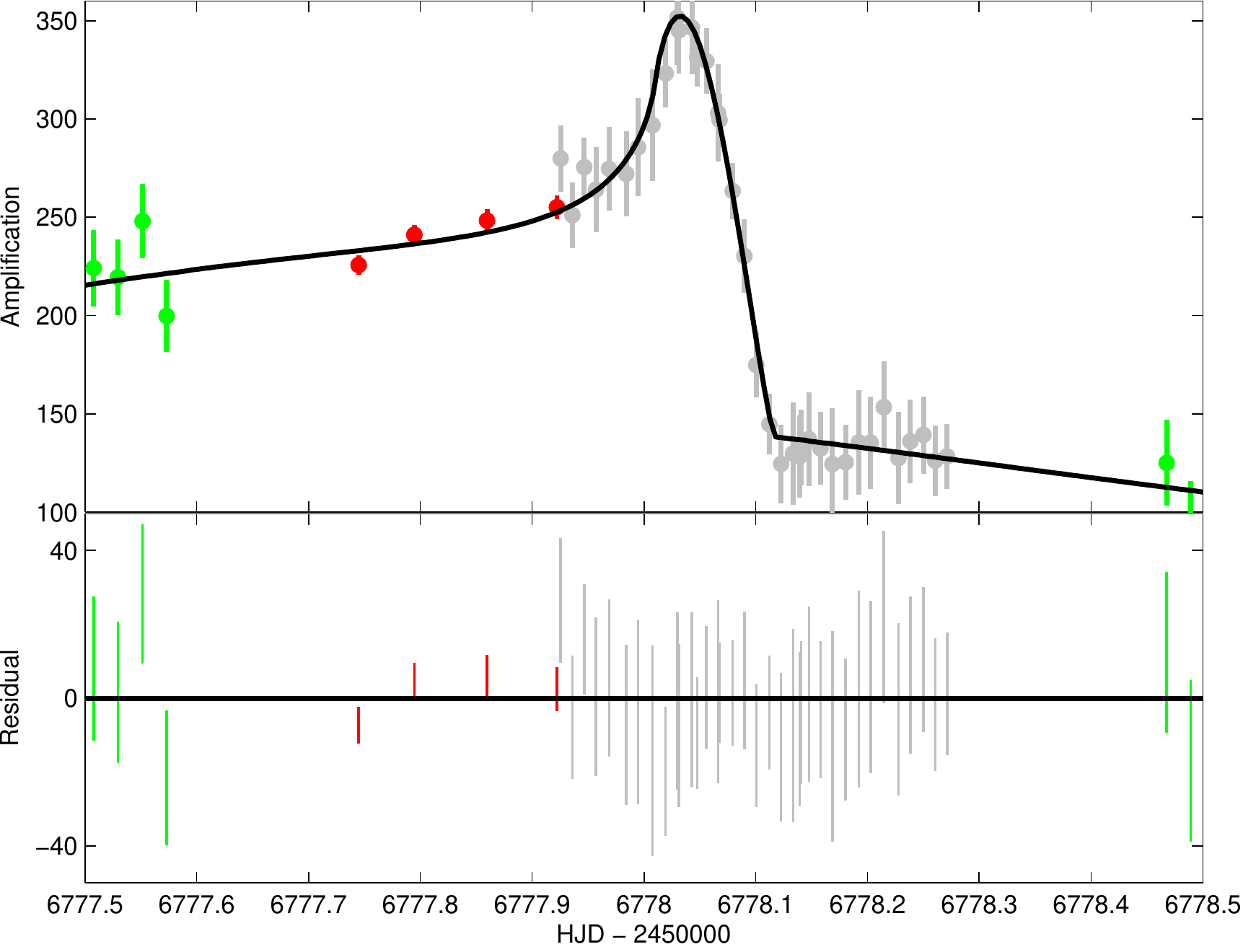}}
 \caption{\label{fig:lightcurve2}This is a close-up view of the light curve for \en as shown in Figure~\ref{fig:lightcurve1}, highlighting the second caustic crossing.}
  \end{figure}

\begin{figure}
\hspace{-25pt}\includegraphics[width=1.3\linewidth]{\FigDir{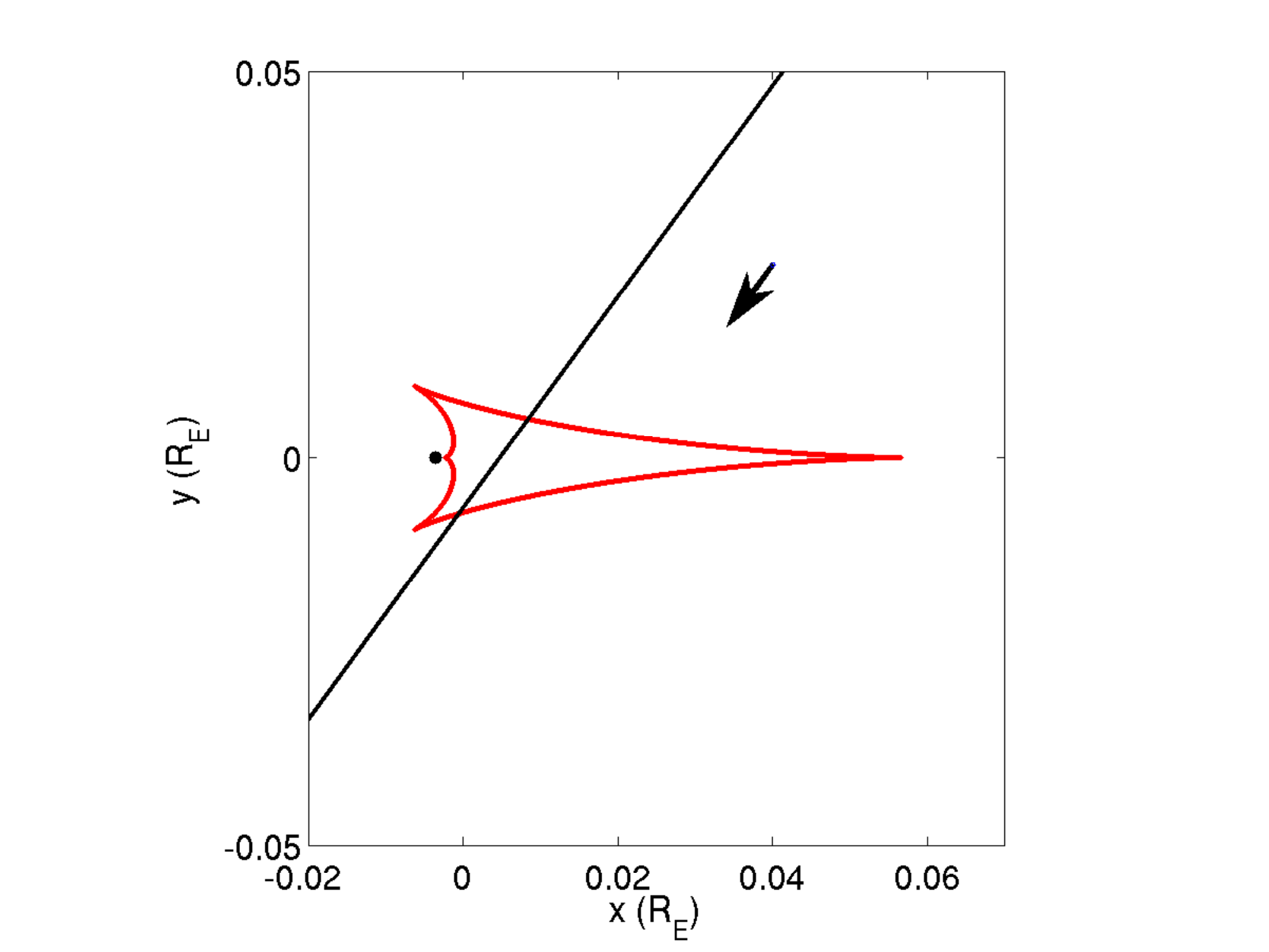}}
 \caption{\label{fig:criticalcaus}The caustic curves (red) corresponding to the best-fitting model for planetary microlensing event \en. The source star track is also shown (black line), and the direction of the source motion is indicated with an arrow.}
  \end{figure}

The parameters for our best-fitting binary lens models are shown in Table~\ref{tab:parameters}. The columns entitled ``Wide'' and ``Close'' list the parameter values for the best-fitting basic binary lens models. The model names --- ``Wide'' and ``Close'' --- reflect the different values of the separation, $s$, between the binary lens masses. In terms of $\chi^{2}$, the Wide model is preferred over the Close model by only $\Delta \chi^{2} \simeq 0.5$. That degenerate models exist with values $s$ and $1/s$  for source star tracks passing close to the central caustic is well-known in planetary microlensing \citep{1999A&A...349..108D}. The light curve data and best-fitting model curve for event \enp is shown in Figures~\ref{fig:lightcurve1} and \ref{fig:lightcurve2}. The caustic curves and source star track is shown in Figure~\ref{fig:criticalcaus}.

\begin{table*}
\begin{center}
\begin{tabular}{lcccc}
\hline
Parameters & Wide & Close & Wide+parallax & Close+parallax\\ \hline
$t_0$ 	&	 6777.335	&	 6777.306	&	 6777.319	&	 6777.309	\\
(JD) 	&	 0.058	&	 0.051	&	 0.045	&	 0.052	\\ \hline
$t_E$ 	&	 116.3	&	 107.3	&	 130.5	&	 126.4	\\
(day) 	&	 12.2	&	 11.4	&	 12.4	&	 10.5	\\ \hline
$u_0 \times 10^3$ 	&	 3.7	&	 4.3	&	 3.6	&	 3.9	\\
 	&	 0.4	&	 0.5	&	 0.4	&	 0.4	\\ \hline
$q\times 10^3$ 	&	 4.78	&	 4.85	&	 4.60	&	 4.31	\\
 	&	 0.13	&	 0.31	&	 0.37	&	 0.07	\\ \hline
$s$ 	&	 1.347	&	 0.760	&	 1.358	&	 0.757	\\
 	&	 0.024	&	 0.014	&	 0.021	&	 0.013	\\ \hline
$\alpha$ 	&	 2.203	&	 2.270	&	 2.229	&	 2.284	\\
(rad) 	&	 0.044	&	 0.045	&	 0.030	&	 0.046	\\ \hline
$\rho\times 10^4$ 	&	 2.78	&	 2.69	&	 2.61	&	 2.43	\\
	&	 0.33	&	 0.31	&	 0.27	&	 0.27	\\ \hline

$\pi_{EN}$ 	&	 ...	&	 ...	&	 -1.81	&	 -2.24	\\
 	&	 ...	&	 ...	&	 0.22	&	 0.51	\\ \hline
$\pi_{EE}$ 	&	 ...	&	 ...	&	 0.06	&	 0.57	\\
 	&	 ...	&	 ...	&	 0.18	&	 0.48	\\ \hline
$\chi^2$ 	&	 12150.2	&	 12150.6	&	 12127.4	&	 12136.7	\\
dof 	&	 12178	&	 12178	&	 12176	&	 12176	\\ \hline
$\chi^2_N$ 	&	 0.99771	&	 0.99775	&	 0.99601	&	 0.99677	\\

\end{tabular}
\end{center}
\caption{\label{tab:parameters}Best-fitting binary lens model parameters for microlensing event \en.
}
\end{table*}

The effect of parallax arising from Earth's orbital motion around the Sun  \citep{1992ApJ...392..442G,1995ApJ...454L.125A} can allow an estimation of the distance to the lens system, and thereby a unique determination of the lens system properties --- in particular the absolute values of the masses of the lens components and the separation between them. In terms of modelling microlensing light curves, we introduce two additional parameters $\pi_{E,E}$ and $\pi_{E,N}$ that are the east and north components of the parallax vector $\vec{\pi_E}$. We can find the mass of the host star and its distance from Earth as follows:
\begin{equation*}
M_L = \thE / (\kappa \pi_E) \quad ;  \quad \dl = \frac{\mathrm{AU}}{\pi_E\thE + \pi_{S}}
\end{equation*}
where \thE is the angular Einstein ring radius, $\pi_S = \mathrm{AU}/\ds$ and $\kappa = 4G/(c^2\mathrm{AU}) = 8.144\, \mas/\Msun$ \citep{2000ApJ...542..785G}. In addition to a measurement of parallax, $\pi_E$, we also need an estimate of the distance to the source star, \ds, and an estimate of the absolute angular source star size which, combined with a modelled value of the source star size, gives us an estimate of \thE.  The value of \tE for the basic binary lens models is $\sim 100$ days, so hopes were high that parallax might be detectable in this event. Disappointingly however, when we included parallax in our modelling for this event we could improve the model fit to the data by only $\Delta \chi^{2} \simeq 23$. While this improvement seems {\it prima facie} significant, there are two sources of concern shedding doubt on this measurement of parallax. Firstly, and perhaps most tellingly, the improvement in $\chi^{2}$ comes from the model fitting  a set of incoherent and noisy MOA data in the baseline, giving rise to the improvement in overall $\chi^{2}$. Secondly, the modelled values of $\pi_{E}$ are unusually higher than those seen in typical parallax events --- casting further doubt on the validity of the parallax models. The high degree of blending in this event, see below, possibly explains why parallax was not observed in this event, similar to event OGLE-2015-BLG-1319 \citep{2016arXiv160602292S}. For these reasons we do not consider that we have a reliable measurement of parallax for event \en. Despite the improvements in $\chi^{2}$ afforded by our best-fitting parallax models, we accept the binary lens models without parallax as our favoured explanation for this event. 

Looking at Figures~\ref{fig:lightcurve1} and \ref{fig:lightcurve2}, we see that the Wide binary lens model fits the data well through the second caustic crossing and exit at times $HJD' \simeq 6778.0$ -- $6778.1$ days. The second caustic crossing was only observed by the MOA collaboration, but the interval between the caustic crossings was well covered by MOA, OGLE and Wise observations. No data were recorded during the first caustic crossing.

\section{Source Star}
\label{sec:sourcestar}
Combined with our modelled values of $\rho$ and $\tE$, an estimate of the angular size of the source star \thS provides an estimate of the angular Einstein ring radius \thE via $\thE = \thS / \rho$. This can provide tighter constraints on the binary lens mass and separation, see e.g. \cite{2014ApJ...780..123S}. We can use published colour-radius relationships to obtain an estimate of \thS, once we have an estimate of the source star colour. 

We start by cross-matching the field stars in the MOA and OGLE-IV photometric catalogues in a 2 arcminute region centred on the co-ordinates of event \en. This gives us the following relationship between instrumental MOA and OGLE-IV colours: $(\Rmoa - \IogleIV) = 0.16709\,(\pm 0.0028) * \VmIogleIV + -3.5593\,(\pm 0.0084)$. We then cross-match OGLE field stars for this event between the calibrated OGLE-III photometric database \citep{2011AcA....61...83S} and the OGLE-IV photometry: 

\begin{align*}
\IogleIII &= (4.684 \pm 0.42)\times 10^{-2} + \IogleIV + \ldots \\
          &\ldots + (-6.1 \pm 1.6)\times 10^{-3}\, \VmIogleIV\\
\VmIogleIII &= (0.9165 \pm 0.0021)\, \VmIogleIV + \dots\\
&\ldots + (0.15747\pm0.0053)
\end{align*}

\begin{figure}
\includegraphics[width=84mm]{\FigDir{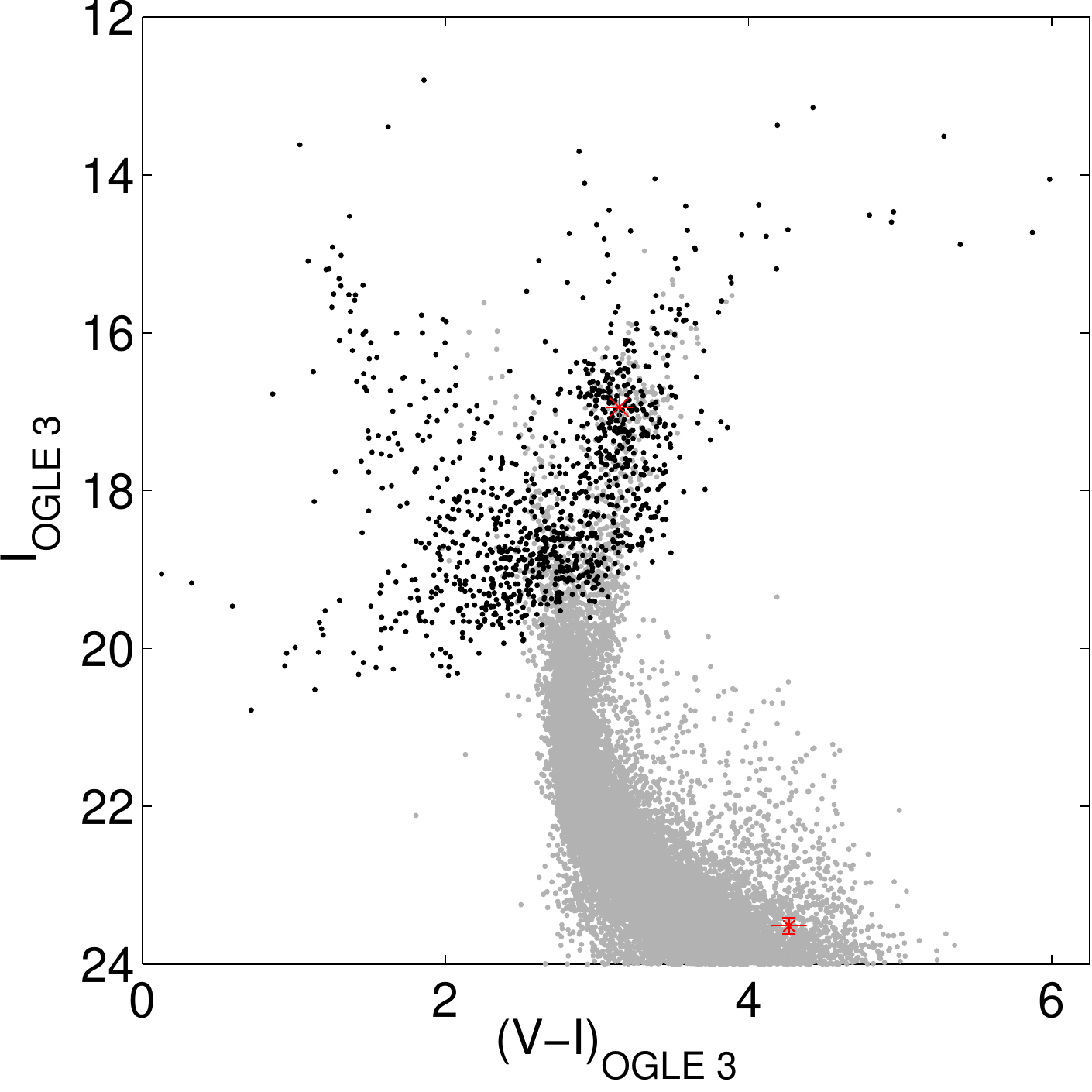}}
 \caption{\label{fig:cmd}Colour magnitude diagram of the OGLE-III field stars within 60 arcsec of event \en (black). The Holtzman (1998) CMD data for Baade's window is shown in grey. The centre of the red clump is indicated (red asterisk) as is the location of the source star.}
  \end{figure}

From the best-fitting model, we obtain the instrumental source magnitude and colour $(\IogleIV, (\Rmoa - \IogleIV))$ and using the above two relationships we obtain the source colour in the calibrated OGLE-III photometric scale:
\begin{align*}
I_{\rm S} & =  23.512 \pm 0.11\\
(V-I)_{\rm S} & =  4.27 \pm 0.11 
\end{align*}
where the errors above include --- in quadrature --- model-dependent errors in estimating instrumental baseline magnitudes, and the uncertainties associated with the linear relationships above. A colour-magnitude diagram of the OGLE-III field stars within 60 arcsec of event \en is shown in Figure~\ref{fig:cmd}.

We need now an estimation of the extinction and reddening suffered by the source star. We make the usual assumption that the source star resides in the Galactic bulge and that it suffers the same degree of extinction and reddening as Red Clump Giant (RCG) stars residing in the bulge. Red clump giant stars are often used as standard candles for Galactic structure studies (e.g. \cite{2013MNRAS.434..595C}). Using a colour magnitude diagram of the OGLE-III field stars for this event we isolate the RCG stars and find the centroid position in the CMD:
\begin{align*}
I_{\rm RCG} & =  16.94 \pm 0.03\\
(V-I)_{\rm RCG} & =  3.15 \pm 0.01.
\end{align*}
We use the following values of the intrinsic luminosity values of Galactic bulge RCGs \citep{2013A&A...549A.147B,2013ApJ...769...88N}: $I_{\rm RCG,0}  =  14.44 \pm 0.04$ with the colour of the red clump being centred at $(V-I)_{\rm RCG,0} =  1.06$ with dispersion $0.06$. Using these intrinsic RCG magnitudes and colours, we estimate the reddening and extinction towards event \enp to be $(A_{\rm I}, E(V-I)) = (2.50, 2.09)\pm (0.05,0.07)$. The intrinsic source colour and magnitude values are therefore $(I, V-I)_{\rm S,0} = (21.02,2.18)\pm (0.12,0.13)$.

In Figure~\ref{fig:cmd} we also show stars observed with the {\it Hubble Space Telescope} towards a region of Baade's window \citep{1998AJ....115.1946H}, where the central position of the red clump for each set of stars have been aligned. Shown on this diagram are the source colour and magnitude values which, with reference to the HST data, suggest that the source is marginally consistent with a star in the bulge. From the relationship given in \cite{2014MNRAS.444..392C} we estimate the effective temperature of the source star to be $\Teff = 3604^{+85} _{-79}$ K. Using this \Teff value, we use the ATLAS stellar atmosphere models of \cite{2000A&A...363.1081C} to obtain the linear limb-darkening coefficients for both the I band data and the MOA-red filter: $(c_{\rm I},c_{R_{\rm MOA}}) = (0.6014, 0.6485)$ where $c_{R_{\rm MOA}}$ is computed as the average of the limb-darkening coefficients in the $R$ and $I$ bands. We apply the stellar radius-colour relationship of \cite{2004A&A...426..297K} to estimate the radius of the source star.  Combined with our modelled value of $\rho = \thS/\thE = (2.78 \pm 0.33) \times 10^{-4}$ we estimate the angular Einstein ring radius to be $\thE = 1.38 \pm 0.43 $ mas.

\section{Lens System Parameter Estimation}
\label{sec:likelihood}
We are unable to set tight constraints on the absolute mass of the host star in the lens system, or its distance from Earth without a credible measurement of microlensing parallax. We resort therefore to a probabilistic analysis of the nature of the lens system \citep{2006Natur.439..437B, 2008ApJ...684..663B}, using a Galactic model \citep{2003ApJ...592..172H} and assuming the distance to the Galactic centre is 8 kpc.

\begin{figure*}
\centering
\begin{varwidth}{0.5\linewidth}  
{\includegraphics[width=8cm]{\FigDir{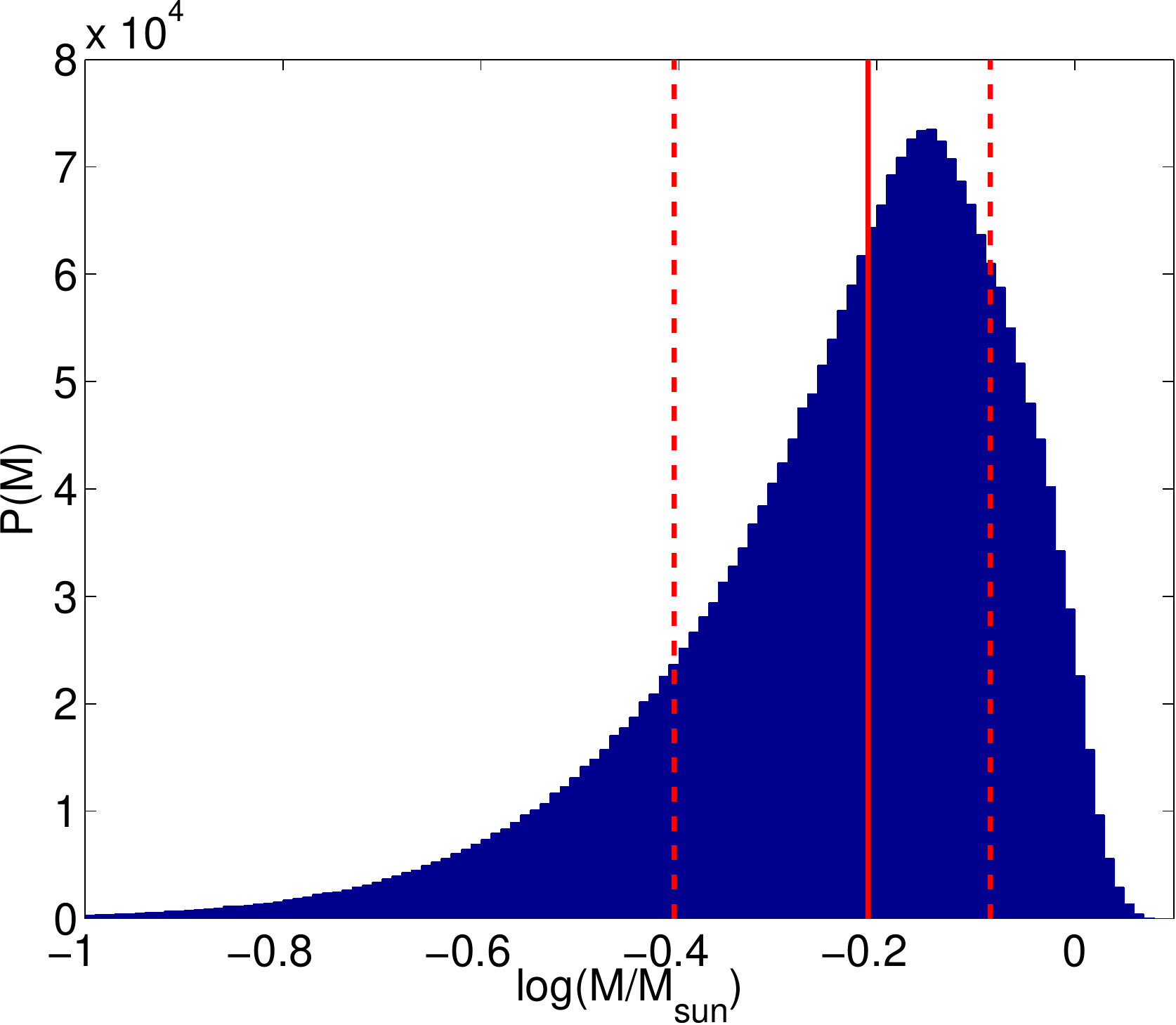}}}
{\includegraphics[width=8cm]{\FigDir{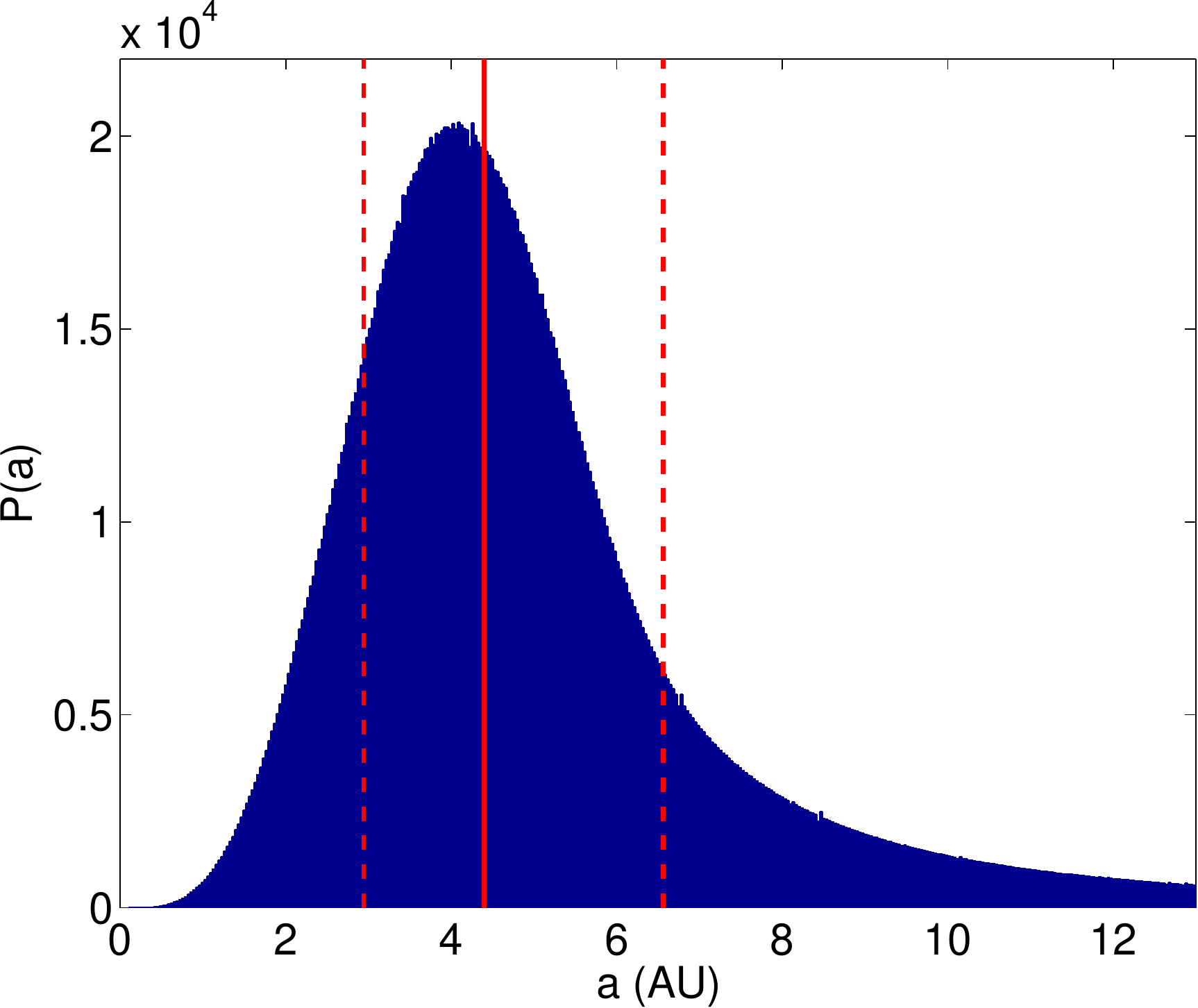}}}\\
\end{varwidth}
\begin{varwidth}{0.5\linewidth}  
{\includegraphics[width=8cm]{\FigDir{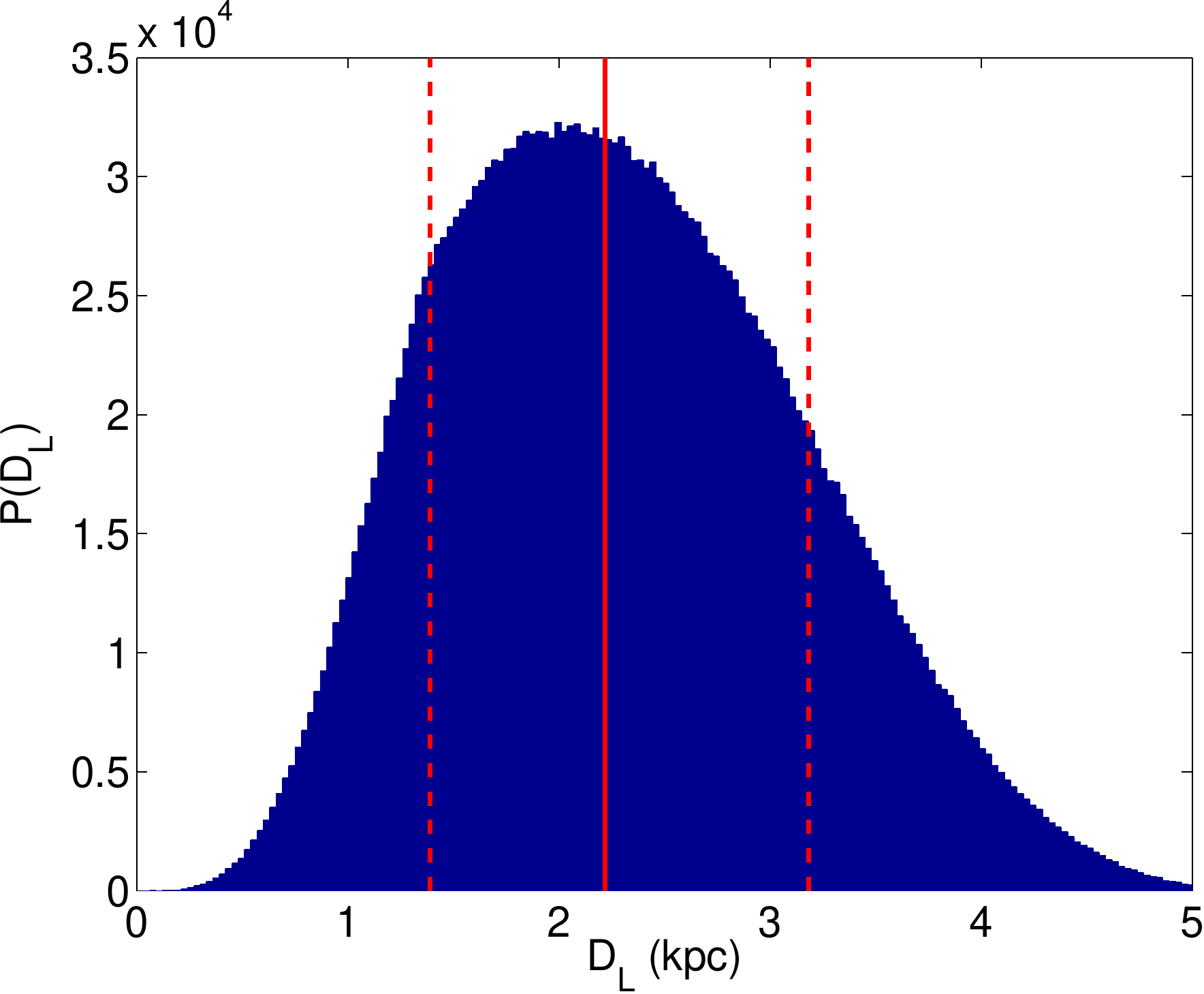}}}\quad\quad
{\includegraphics[width=8cm]{\FigDir{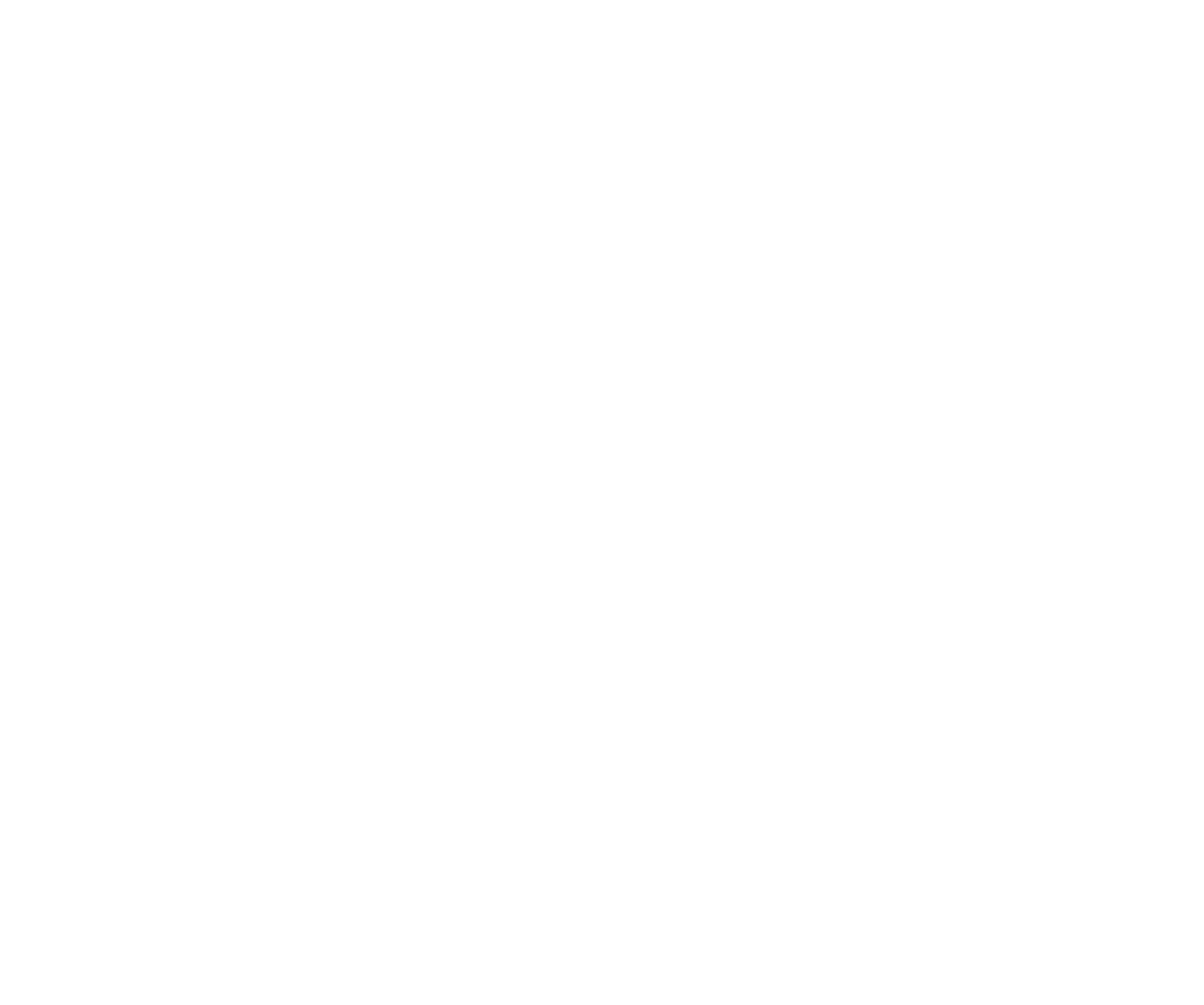}}}\\
\end{varwidth}
\caption{
Lens system parameters from the probability density analysis. Parameter distributions are shown for, clockwise from top left, host star mass, $M$, host star distance \dl and deprojected planet orbital radius, $a$.
Median parameter values and 68\% confidence intervals are shown as vertical red solid and dashed lines, respectively. } 
\label{fig:thetaE_noremnant}
\end{figure*}

Using our observed value of \thE, \tE, and the de-extincted blend flux derived from the OGLE-I band data as the upper limit of the light contributed by the lens in this event, $I_{\mathrm b,0} = 17.202 \pm 0.005$, we find the lens system comprises a \planetmassdashNoRem planet orbiting a \lensmassdashNoRem host star at a deprojected orbital distance of \planetorbitdashNoRem. The distance to the lens system is \lensdistancedashNoRem. The parameter distributions are shown in Figure~\ref{fig:thetaE_noremnant}.

The lens-source proper motion is estimated as $\mu_{\textrm rel} = \thE / \tE = 4.33 \pm 1.43$ mas/yr. Following the estimation of \cite{2007ApJ...660..781B} of when the HST could have resolved the lens and source components for event OGLE-2005-BLG-169L, and considering the estimated lens-source relative proper motion for \en, HST could resolve the lens and source for this event in around 3.7 years following the event peak --- corresponding to the end of 2017. In our Bayesian analysis, most of the light in this event was contributed by the lens system and thus we needn't wait for the separation of the lens and source to allow a measurement of the lens flux by HST. Adaptive optics measurements from the ground will be able to make measurements of the lens flux now for comparison to our Bayesian estimates.
\begin{table}
\begin{center}
\begin{tabular}{llll}
\hline
Physical parameter & Value \\
\hline
Planet mass & \planetmassdashNoRem \\[10pt]
Orbital radius & \planetorbitdashNoRem \\[10pt]
Host mass  & \lensmassdashNoRem \\[10pt]
Lens distance & \lensdistancedashNoRem\\[10pt]
Einstein ring radius & $1.38 \pm 0.43 $ mas\\
\hline
\end{tabular}
\end{center}
\caption{\label{tab:physicalparams}Physical parameters for the favoured binary lens model for microlensing event \en.}
\end{table}

There are a number of important assumptions made in the derivation of lens system parameters via such a Bayesean probability density analysis and the reader is directed to  \cite{2015MNRAS.454..946R} for these.

\section{Discussion and Conclusion}
\label{sec:conclusion}
We found the anomalous signal in gravitational microlensing event \en to be consistent with a planetary lens system. The source star passed through the central caustic, with the second caustic crossing being well recorded by the MOA microlensing survey collaboration. Observations at epochs between the unrecorded first caustic crossing and the second caustic crossing were made by the OGLE, Wise and MOA collaborations. Further observations were made by the RoboNET/LCOGT and MiNDSTEp groups, but these were made at times away from the planetary signal. The best-fitting binary lens model for this event has a binary lens mass ratio of $(4.78 \pm 0.13)\times 10^{-3}$ and the binary lens components are separated by $s = 1.347 \pm 0.024$, with a degenerate solution having $s= 0.760 \pm 0.014$. This lens mass ratio is a few times greater than that found in the earlier work by \cite{2016MNRAS.457.4089S}, in which the mass ratio for this event was reported to be $q\sim 1.4\times10^{-3}$. This lower mass ratio  was obtained from a coarse search of the binary lens microlensing parameter space, as this event was only one of 29 anomalous events analysed in that work.

A believable measurement of parallax for this event was not possible. We did, however, estimate the angular radius of the source star and thereby estimate the angular size of the Einstein ring radius, \thE. The source star is rather faint, and very red. There is a possibility that the source may be blended with a nearby red star, causing an incorrect identification of the source star type. 

A Bayesian probability analysis for the lens system parameters for this event suggests that the binary lens system for \en is consistent with a \planetmassdashNoRem planet orbiting a \lensmassdashNoRem host star at a deprojected orbital separation of \planetorbitdashNoRem. The distance to the lens system is \lensdistancedashNoRem. This measurement of the lens mass is consistent with a K-dwarf star. 
In our Bayesian analysis we assumed that most of the light from this event is contributed by the lens star and thus lens flux measurements with an adaptive-optics system will provide a check on our Bayesian estimates of the lens system.

In any event, planet \enp can be added to the growing list of planets discovered by microlensing against which planetary formation theories can be tested.

\section{Acknowledgements}
NJR is a Royal Society of New Zealand Rutherford Discovery Fellow. AS is a University of Auckland Doctoral Scholar. TS acknowledges financial support from the Japan Society for the Promotion of Science (JSPS) under grant numbers JSPS23103002, JSPS24253004 and JSPS26247023. NK is supported by Grant-in-Aid for JSPS Fellows. The MOA project is supported by JSPS grants JSPS25103508 and JSPS23340064 and by the Royal Society of New Zealand Marsden Grant MAU1104.  NJR acknowledges the contribution of NeSI high-performance computing facilities to the results of this research. NZ's national facilities are provided by the NZ eScience Infrastructure and funded jointly by NeSI's collaborator institutions and through the Ministry of Business, Innovation \& Employment's Research Infrastructure programme (https://www.nesi.org.nz).

The OGLE team thanks Profs. M.~Kubiak and G.~Pietrzy{\'n}ski, former members of the OGLE team, for their contribution to the collection of the OGLE photometric data over the past years. The OGLE project has received funding from the National Science Centre, Poland, grant MAESTRO 2014/14/A/ST9/00121 to AU.

YS is a NASA Postdoctoral Program Fellow.

The Danish 1.54 m telescope is operated based on a grant from the Danish Natural Science Foundation (FNU). The MiNDSTEp monitoring campaign is powered by ARTEMiS (Automated Terrestrial Exoplanet Microlensing Search) \citep{2008AN....329..248D}. This publication was made possible by NPRP grants \# X-019-1-006 and 09-467-1-078 from the Qatar National Research Fund (a member of Qatar Foundation). KH acknowledges support
from STFC grant ST/M001296/1. GD acknowledges Regione Campania for support from POR-FSE Campania 2014-2020. 
TCH acknowledges support from the Korea Research Council of Fundamental Science \& Technology (KRCF) via the KRCF Young Scientist Research Fellowship Programme and for financial support from KASI travel grant number 2014-1-400-06. 
J. Surdej acknowledges support from the Communaut\'{e} fran\c{c}aise de Belgique - Actions de recherche concert\'{e}es - Acad\'{e}mie Wallonie-Europe. This work has made extensive use of the ADS service, for which we are thankful.

This work makes use of observations from the LCOGT network, which includes three SUPAscopes owned by the University of St Andrews. The RoboNet programme is an LCOGT Key Project using time allocations from the University of St Andrews, LCOGT and the University of Heidelberg together with time on the Liverpool Telescope through the Science and Technology Facilities Council (STFC), UK. This research has made use of the LCOGT Archive, which is operated by the California Institute of Technology, under contract with the Las Cumbres Observatory.

The authors thank Matthew Penny for his comments on this manuscript.

\vspace{0.25cm} 

\noindent $^{1}$Department of Physics, University of Auckland, Private Bag 92019, Auckland, New Zealand\\
$^{2}$Department of Physics, University of Notre Dame, Notre Dame, IN 46556, USA\\
$^{3}$Department of Earth and Space Science, Graduate School of Science, Osaka University, 1-1 Machikaneyama, Toyonaka, Osaka 560-0043, Japan\\
$^{4}$Institute of Natural and Mathematical Sciences, Massey University, Private Bag 102-904, North Shore Mail Centre, Auckland, New Zealand\\
$^{5}$Warsaw University Observatory, Al.~Ujazdowskie~4, 00-478~Warszawa, Poland\\
$^{6}$Jet Propulsion Laboratory, California Institute of Technology, 4800 Oak Grove Drive, Pasadena, CA~91109, USA\\
$^{7}$School of Physics and Astronomy, Tel-Aviv University, Tel-Aviv 69978, Israel\\
$^{8}$Institute and Centre for Star and Planet Formation, University of Copenhagen, {\O}ster Voldgade 5, 1350 Copenhagen K, Denmark\\
$^{9}$SUPA, School of Physics and Astronomy, University of St. Andrews, North Haugh, St Andrews, KY16 9SS, United Kingdom\\
$^{10}$Las Cumbres Observatory Global Telescope Network, Inc., 6740 Cortona Drive, Suite 102, Goleta, CA  93117, USA\\
$^{11}$Astronomisches Rechen-Institut, Zentrum f{\"u}r Astronomie der Universit{\"a}t,  Heidelberg, M{\"o}nchhofstr. 12-14, 69120 Heidelberg, Germany\\
$^{12}$Solar-Terrestrial Environment Laboratory, Nagoya University, Nagoya, 464-8601, Japan\\
$^{13}$Astrophysics Science Division, NASA Goddard Space Flight Center, Greenbelt, MD 20771, USA\\
$^{14}$School of Physics, The University of New South Wales, Sydney NSW 2052, Australia\\
$^{15}$Okayama Astrophysical Observatory, National Astronomical Observatory, 3037-5 Honjo, Kamogata, Asakuchi, Okayama 719-0232, Japan\\
$^{16}$Nagano National College of Technology, Nagano 381-8550, Japan\\
$^{17}$Tokyo Metropolitan College of Industrial Technology, Tokyo 116-8523, Japan\\
$^{18}$School of Chemical and Physical Sciences, Victoria University, Wellington, New Zealand\\
$^{19}$Mt. John Observatory, P.O. Box 56, Lake Tekapo 8770, New Zealand\\
$^{20}$Department of Physics, Faculty of Science, Kyoto Sangyo University, 603-8555 Kyoto, Japan\\
$^{21}$Department of Astronomy, Ohio State University, 140 W.~18th Ave., Columbus, OH~43210, USA\\
$^{22}$Department of Physics, University of Warwick, Gibbet Hill Road, Coventry, CV4~7AL,~United~Kingdom\\
$^{23}$Qatar Environment and Energy Research Institute (QEERI), HBKU, Qatar Foundation, Doha, Qatar\\
$^{24}$Boston University, Boston MA, USA\\
$^{25}$Dipartimento di Fisica ``E. R. Caianiello'', Universit\`a di Salerno, Via Giovanni Paolo II, 84084 Fisciano (SA), Italy\\
$^{26}$Istituto Nazionale di Fisica Nucleare, Sezione di Napoli, Napoli, Italy\\
$^{27}$IPAC, Mail Code 100-22, Caltech, 1200 E. California Blvd., Pasadena, CA 91125\\
$^{28}$Institut d'Astrophysique et de G\'eophysique, Universit\'e de Li\`ege, All\'ee du 6 Ao\^ut, B\^at. B5c, 4000 Li\`ege, Belgium\\
$^{29}$Georg-August-Universi{\"a}t, Altes Rathaus, Markt 9, 37073 G{\"o}ttingen, Germany\\
$^{30}$Korea Astronomy and Space Science Institute, Daejeon 305-348, Republic of Korea\\
$^{31}$European Southern Observatory, Karl-Schwarzschild-Str. 2, 85748 Garching bei M{\"u}nchen, Germany\\
$^{32}$Space Telescope Institute, 3700 San Martin Drive, Baltimore, MD 21218, USA\\
$^{33}$Jodrell Bank Centre for Astrophysics, The University of Manchester, Manchester, M13 9PL, United Kingdom\\
$^{34}$Max Planck Institute for Astronomy, K\"onigstuhl 17, 69117 Heidelberg, Germany\\
$^{35}$Perth Observatory, 337 Walnut Rd, Bickley WA 6076, Australia\\
$^{36}$Instituto de Astrof\'isica, Facultad de F\'isica, Pontificia Universidad Cat\'olica de Chile, Av. Vicu\~na Mackenna 4860, 7820436 Macul, Santiago, Chile\\
$^{37}$Department of Physics, Sharif University of Technology, P. O. Box 11155-9161 Tehran, Iran\\
$^{38}$Istituto Internazionale per gli Alti Studi Scientifici (IIASS), 84019 Vietri Sul Mare (SA), Italy\\
$^{39}$Max-Planck-Institute for Solar System Research, Justus-von-Liebig-Weg 3, 37077 G\"ottingen, Germany\\
$^{40}$Astrophysics Group, Keele University, Staffordshire, ST5 5BG, United Kingdom\\
$^{41}$South African Astronomical Observatory, PO Box 9, Observatory 7935, South Africa\\
$^{42}$Astrophysics Research Institute, Liverpool John Moores University, IC2, Liverpool Science Park, 146 Brownlow Hill, Liverpool L3~5RF, United~Kingdom\\
$^{A}$Microlensing Observations in Astrophysics (MOA)\\
$^{B}$Optical Gravitational Lensing Experiment (OGLE)\\
$^{C}$Wise Observatory Group\\
$^{D}$Microlensing Network for the Detection of Small Terrestrial Exoplanets (MiNDSTEp)\\
$^{E}$RoboNET\\

\bibliographystyle{latest}

\bibliography{microlensing}

\end{document}